%% file: OffPath.tex
\documentclass{sig-alternate}

\usepackage{graphicx}
\usepackage{epstopdf}%

\makeatletter
\def\ps@headings{%
\def\@oddhead{\mbox{}\scriptsize\rightmark \hfil \thepage}%
\def\@evenhead{\scriptsize\thepage \hfil \leftmark\mbox{}}%
\def\@oddfoot{}%
\def\@evenfoot{}}
\makeatother
\usepackage{amsmath}%
\usepackage{amsfonts}%
\usepackage{amssymb}%
\usepackage{graphicx}
\usepackage{wrapfig}
\usepackage{subfigure}
\usepackage[all]{xy}
\usepackage{sidecap}
\usepackage{url}

\newcommand{\ignore}[1]{}

\newcommand{\fullppr}[1]{}

\newcommand{\norfc}[1]{RFC~#1}

\ignore{

\newcommand{\qed}{\nobreak \ifvmode \relax \else
      \ifdim\lastskip<1.5em \hskip-\lastskip
      \hskip1.5em plus0em minus0.5em \fi \nobreak
      \vrule height0.75em width0.5em depth0.25em\fi}
}

\newcommand{\mal}{Oscar}

\newcommand{\wndsize}{\mathrm{\left|\textit{wnd}\right|}}
\newcommand{\wnd}{$\mathrm{\textit{wnd}}$}

\begin{document}

\title{Off-Path Hacking:\\\Large{\em The Illusion of Challenge-Response Authentication}}
\numberofauthors{1}
\author{
\alignauthor
Yossi Gilad\titlenote{mail@yossigilad.com} and Amir Herzberg\titlenote{amir.herzberg@gmail.com} and Haya Shulman\titlenote{haya.shulman@gmail.com}\\
\affaddr{~}\\
\affaddr{Computer Science Department\\Bar Ilan University\\Ramat Gan, Israel}
}

\maketitle

\input{abstract}
\input{intro}
\input{dns-poisoning}
\input{TCP}
\input{Exploits}

\input{conclusions}

\bibliographystyle{abbrv}
\bibliography{../../bib/NetSec,../../bib/rfc}

\end{document}

%% file: abstract.tex
\begin{abstract}
Everyone is concerned about the Internet security, yet most traffic is not cryptographically protected. The usual justification is that most attackers are only {\em off-path} and cannot intercept  traffic; hence, challenge-response mechanisms suffice to ensure authenticity. Usually, the challenges re-use existing `unpredictable'  header fields to protect  widely-deployed protocols such as TCP and DNS. 

We argue that this practice may often only give an {\em illusion of security}. We present recent off-path TCP injection and DNS poisoning attacks, enabling attackers to circumvent existing challenge-response defenses. 
Both TCP and DNS attacks are non-trivial, yet very efficient and practical. The attacks foil widely deployed security mechanisms, such as the Same Origin Policy, and allow 
a wide range of exploits, e.g., long-term caching of malicious objects and scripts. 

We hope that this article will motivate adoption of cryptographic mechanisms such as SSL/TLS, IPsec and DNSSEC, and of correct, secure challenge-response mechanisms.   
\end{abstract}


\ignore{
\begin{abstract}
Even a weak, off-path attacker can efficiently attack basic Internet infrastructure and protocols. 
Specifically, contrary to common belief, we show how off-path attackers can poison DNS and inject data into TCP connections. 
The attacks are non-trivial, yet very efficient and practical, and allow circumvention of much of the currently-deployed security mechanisms, such as Same Origin Policy (SOP). The attacks motivate adoption of cryptographic mechanisms such as SSL/TLS and DNSSEC, as well as improved analysis of vulnerabilities of Internet protocols.  
\end{abstract}
}

%% file: intro.tex
\section{Introduction}
\begin{figure} [t]
  \begin{center}
    \includegraphics[width=0.4\textwidth]{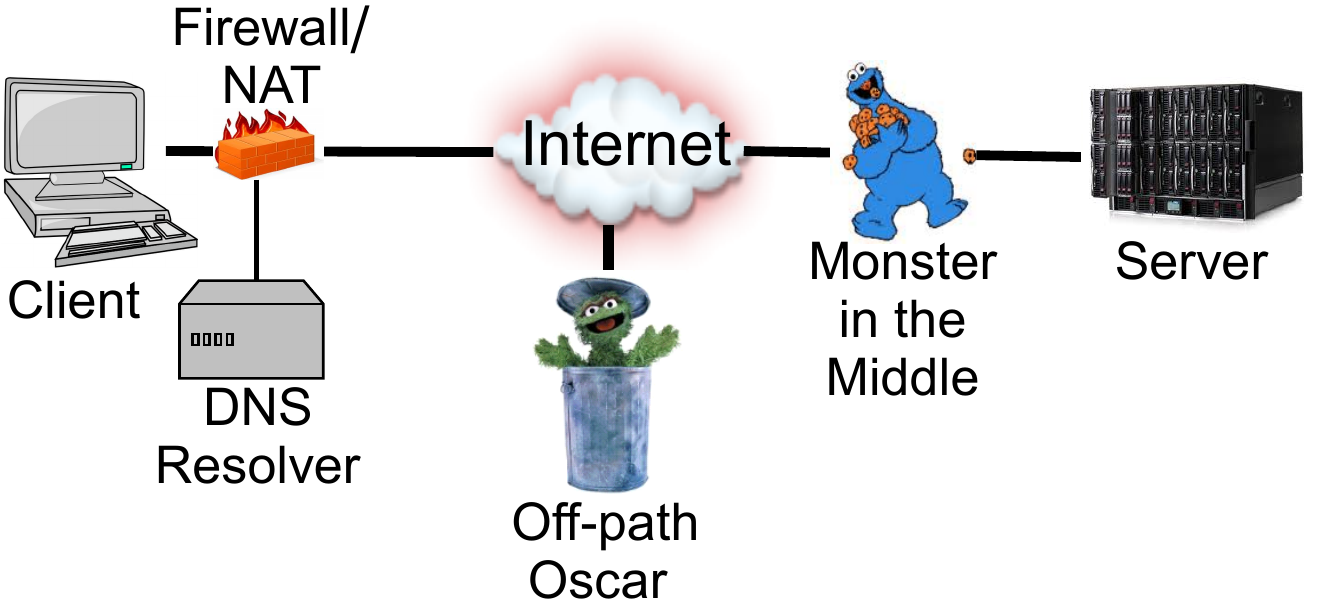}
  \end{center}
  \caption{Off-Path Attacker Network Model.}
   \label{fig:ourmodel}
\end{figure}

Since 1989 \cite{Bellovin:security:problems:in:TCP}, experts have been arguing that Internet security requires cryptographic protocols, ensuring security against {\em Monster-in-the-Middle (MitM) attackers}. A MitM attacker is located on the path of the communicating parties, and can manipulate the communication between them in any way, i.e., intercept, modify, block and inject spoofed packets; see the {\em MitM Cookie Monster} in Figure~\ref{fig:ourmodel}. 

The information security community invested huge efforts in developing cryptographic schemes and protocols, standards and products, providing security against MitM attackers, such as IPsec, SSL/TLS, Secure-BGP and DNSSEC. In spite of all these efforts, and although Internet security is well recognised to be critical, most Internet traffic is still not cryptographically protected. 
For example, we 
found that only about $6\%$ of the TCP traffic is cryptographically protected with SSL/TLS (based on CAIDA dataset of $3$ million packets \cite{CAIDATraces}); and less than 1\% of the DNS resolvers enforce DNSSEC (cryptographic) validation \cite{olafur:nat:res}. 

We believe that the main reason for the underutilisation of cryptography, is an {\em illusion of security} against network-based attacks, due to {\em two false beliefs}. The first false belief, is that in reality, attackers can rarely obtain MitM capabilities, and even when they can, they are reluctant to do so since such activities may lead to detection. We claim that this is incorrect; there are common scenarios where attackers may obtain MitM capabilities, e.g., by accessing wireless communication, by manipulations of the (largely unprotected) routing mechanisms, or by controlling some intermediate device; furthermore, such attacks are often carried out, without detection and repercussions, e.g., BGP route hijacking happens frequently \cite{ballani2007study}. 

However, in this article, we focus on the second false belief, which is that current, non-cryptographic, Internet protocols already provide sufficient protection against {\em typical, common} attackers, and in particular, against {\em off-path attackers}. 
Unlike a MitM attacker, an off-path attacker cannot observe or modify legitimate packets sent between other parties, however, it can transmit packets with a spoofed (fake) source IP address - impersonating some legitimate party, as illustrated by {\em Off-path Oscar} in Figure~\ref{fig:ourmodel}. 

Our main goal in this paper is to convince that this second belief is (also) false, and that {\em current Internet protocols are often vulnerable even to an off-path attacker}. Specifically, we discuss few recent results, that allow off-path attacks on basic Internet protocols: traffic injection into TCP connections and DNS cache poisoning.

Note that spoofed packets are used in many attacks, most notably, in Denial of Service (DoS) attacks. 
Significant efforts are made to make spoofing less readily available to attackers, most notably ingress filtering (RFC 3704). However, IP spoofing is still possible via many ISPs, 
see \cite{SpooferProject,conf/imc/BeverlyBHc09}. 

The key, to the off-path attacks that we discuss, is {\em circumvention of challenge-response defenses}, which are often relied upon to distinguish between (spoofed) packets from an off-path attacker and (legitimate) packets from legitimate communication end point. In order to authenticate a response from a server, a client sends a random challenge with the request, which is returned with the response. Since an off-path attacker, which we dub Oscar, cannot eavesdrop on packets exchanged between the server and the client, it appears that he would have to guess the challenge; hence, the (sufficiently long, random) challenge allows to prevent Oscar from crafting a packet with a valid challenge.
The security of most Internet applications, e.g., email, web surfing, and most peer-to-peer applications, relies on challenge-response mechanisms, mostly as part of 
the underlying TCP and DNS protocols. For example, the widely-used web-security mechanisms based on cookies and on the `same origin policy' in general, depend on the security of both TCP and DNS. 

Note that, trivially, {\em challenge-response mechanisms are ineffective against MitM attackers}, since they are able to eavesdrop on the challenges. The false sense of security is due to the two false beliefs mentioned above: that MitM capabilities are `rarely practical' and that existing challenge-response mechanisms, in particular, in TCP and DNS, provide sufficient defenses against the (weaker and common) off-path attackers.  
The belief that off-path attackers cannot {\em inject} traffic into a TCP connection is even stated in RFCs and standards, e.g., \norfc{4953}, discussing TCP spoofing attacks. The reasoning is that TCP specifications and implementations were enhanced to provide security against such adversaries, who are incapable of eavesdropping to communication: modern TCP implementations randomise the 32-bit sequence number~\cite{rfc6528}, and many also randomise the 16-bit client port~\cite{rfc6056}. To successfully inject data into a TCP stream, the attacker must provide valid values in both fields. Indeed, since its early days, most Internet traffic is carried over TCP - and is not cryptographically protected, in spite of warnings~\cite{Bellovin:security:problems:in:TCP,Bellovin:Look:Back:at:TCP:IP:Security,Morris85}. 

In contrary to this second belief, we argue that {\em there are (common) situations allowing  off-path attacks}, i.e., where Oscar can circumvent challenge-response mechanisms. One approach \cite{frag:vulnerable:journal,DBLP:journals/corr/abs-1205-4011}
exploits {\em IP fragmentation}, i.e., division of IP packets into several fragments where only one fragment contains the response-field; this allows off-path attacks that block, modify and even intercept fragments or whole packets in some scenarios. 

However, in this article we focus on another approach: in order to circumvent deployed challenge-response mechanisms in TCP and DNS, we exploit the fact that to ease deployment, challenge-response defenses for these protocols mostly or wholly {\em reuse existing fields}; i.e., challenges are fields which already exist in requests and are echoed in responses for some other purpose. In TCP, the `challenge'~fields are only reused fields: sequence number and client port; in DNS, there is a short `dedicated' challenge~field (16~bits), but since it is insufficient to foil attacks, other fields are `reused' as challenges, including source port, source/destination IP addresses, and the query itself, see \norfc{5452}. 

The use of a randomised (16-bit) source port field, that maps responses to the client process which issued the request, is a 
widely used `best practice' against off-path attacks; see RFCs 6056, 5452 and \cite{kaminsky:dns}. 

We argue that such {\em dual-use} of an existing field for challenge-response, while conveniently allowing deployment of defenses only on the client side without requiring coordinated adoption by the server, is often vulnerable. 
Specifically, we discuss attacks that allow an off-path attacker to {\em learn the source port} and other `dual-use' fields. This allows off-path TCP injection and DNS cache poisoning.



\subsection{History of Off-Path Attacks} 
TCP and DNS are basic protocols, and off-path attacks on their authenticity - TCP injection and DNS poisoning - impact almost all Internet applications. As such, it is a common belief that they ensure integrity against an off-path attacker. However, security against off-path (or MitM) attackers was not of the original design goals of these protocols, and only minimal changes were done to the specifications to support challenge-response defenses, e.g., selecting identifiers at random. 

\begin{figure*}[t]
  \begin{center}
    \includegraphics[width=\textwidth]{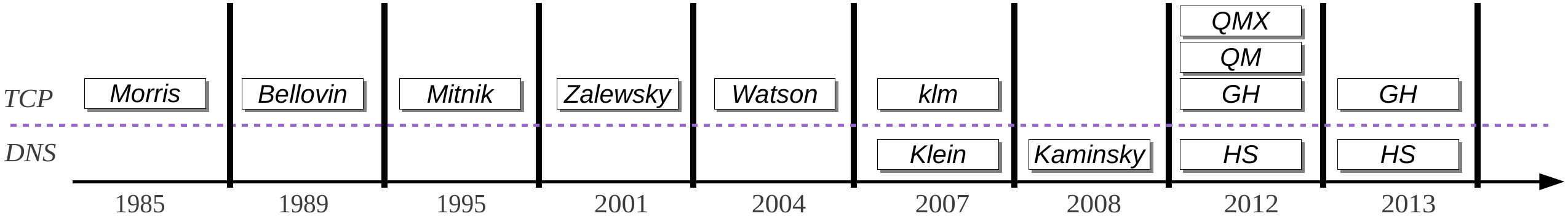}
  \end{center}
  \caption{Time-line of DNS Poisoning and TCP Injection attacks.}
   \label{fig:attacks}
\end{figure*}

Indeed, over the years, significant attention and efforts were dedicated to validating and improving the off-path security of TCP and DNS - and numerous off-path attacks were launched, some of them widely publicised. 
In Figure~\ref{fig:attacks} we present a `time-line' of important attacks and security improvements, for both TCP (upper row) and DNS~(lower~row). 

The time-line begins in 1985 with Morris publication of TCP injection attack based on the use of predictable sequence numbers~\cite{Morris85}, and Bellovin's seminal paper from 1989~\cite{Bellovin:security:problems:in:TCP}, pointing out that security should {\em not} be based on the presumed off-path protection of DNS and TCP. Bellovin presented vulnerabilities of (some) TCP implementations to off-path attacks, and discussed potential exploits and defenses.

Unfortunately, in spite of these warnings, until 1995 most TCP stacks still used trivially-predictable {\em initial sequence number (ISN)}. This changed only after the infamous attack by Mitnick on Shimomura \cite{Shimomura:1995:TPC:546748}, that utilized a TCP injection.  After the attack, many implementations changed to `less predictable' ISN choices. However, in 2001, Zalewski showed that most implementations are still `sufficiently predictable', allowing off-path attacks; this motivated adoption of more random choice of ISNs in most operating systems, as standardised in \norfc{6528}. 


One additional TCP injection attack was presented in~2004 by Watson \cite{watson2004slipping}. This attack only injected a `RST' packet, breaking up a connection, and focused on long-lived connections using known client (and server) ports and addresses, as used at the time, in particular, by the Internet routing protocol BGP. To address this concern, many TCP implementations began also using `unpredictable' client ports. 

In 2007, there were two surprising results: (1) a TCP injection attack presented by the pseudonym author {\em klm} in Phrack magazine \cite{klm:phrack:07}, and (2) a DNS poisoning attack exploiting poor random-number generators \cite{Klein07:OpenBSD}. Both attacks were clever and significant although with limited scope. In particular, the attack on TCP worked only against Windows machines, connected directly to the Internet (rather than via firewall, as usually is the case), and did not handle concurrent connections.   

Kaminsky presented an even more significant DNS poisoning attack in 2008, which allowed efficient off-path poisoning of most DNS resolver implementations at the time~\cite{kaminsky:dns} (see Section \ref{sc:dns}). 
The response to this attack was rapid adoption of additional `patches', mostly, more challenge-response fields, increasing the randomisation and therefore (hopefully) making the attack impractical; the most notable patch was {\em source port randomisation (SPR)}; see \norfc{5452}. 

Following 2008, there were several years without additional off-path attacks; Kaminsky's attack was addressed by SPR and other `patches' to resolvers, and klm's attack was impractical and not widely known (due to the unusual venue). This changed dramatically in 2011-2013, with the publication of {\em eight} new off-path attacks. The first, in 2011, was an attack on fragmented IP traffic \cite{frag:vulnerable:journal}. This was followed, by {\em two} new DNS poisoning attacks (in 2012 \cite{DBLP:conf/esorics/HerzbergS12} and 2013 \cite{DBLP:journals/corr/abs-1205-4011}), a connection-exposing attack \cite{conf/pet/GiladH12} and {\em four} TCP injection attacks \cite{woottcp,wwwtcp,snptcp,CCS12:tcp}. We discuss some of these attacks in this paper. 

\subsection{Malicious Agents} \label{intro:agents}
Some of the off-path attacks require, in addition to the spoofing ability, also a {\em  malicious agent} in the victim's network or host. We now explain the different agent models.

A {\em user-level zombie} is a machine controlled by the adversary, e.g., compromised by malware, in the victim's network. 
There appears to be a significant amount of attacker controlled hosts (zombies) on the Internet. 

A  {\em puppet}  \cite{journals/tissec/AntonatosALA08} is weaker agent: a restricted malicious script or applet running in web-browser sandbox. Attacks relying on a puppet agent require (only) that a client in the victim network `surfs' to the attacker's web-site, enabling the adversary to run such a script.
The script is restricted by {\em same origin policy} (described in RFC 6454), and can only communicate via the browser, i.e., request (and receive) HTTP objects (no access to TCP/IP packet headers). 

Puppets are usually easier to obtain and control compared to {\em  zombies}, since browsers normally run scripts automatically upon opening a web-site, while zombies require installation (of malware).

%% file: dns-poisoning.tex
\section{DNS Cache Poisoning}\label{sc:dns}\label{sc:spr}
Challenge-response mechanisms allow DNS resolvers to authenticate legitimate DNS responses, thus preventing cache poisoning by off-path attackers. Until the year~2008 the only (widely deployed) challenge-response mechanism was the 16-bit {\em transaction identifier} (TXID). 
\subsection{Kaminsky's DNS Cache Poisoning}
\begin{figure}[!ht]
\centering
	\begin{center}
			    \epsfig{scale=0.47,file=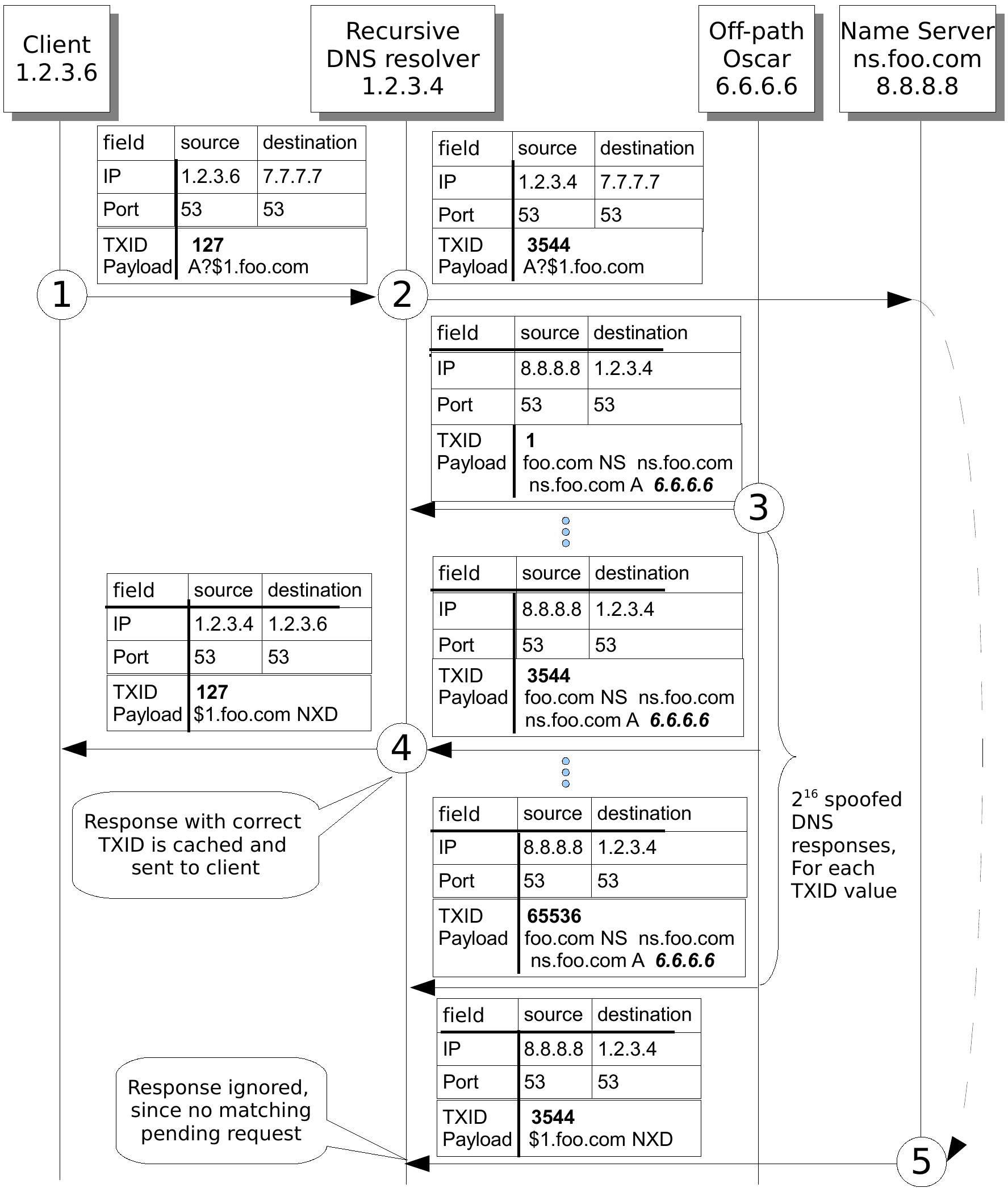}
	\end{center}	
\caption{Kaminsky's Attack.}
 \label{fig:kaminsky}
\end{figure}

In 2008, Kaminsky~\cite{kaminsky:dns} presented an efficient cache poisoning attack against resolvers which authenticated responses using only the TXID. We briefly describe this attack. The attack assumes a known source port, and for concreteness we used source port 53. The steps of the attack, illustrated in Figure~\ref{fig:kaminsky}, are the following: 

(1) the attacker triggers a DNS request for a random subdomain of the victim domain {\sf 1\$.foo.com}. There are different techniques to trigger DNS requests, e.g., a benign client visits a web page of the attacker; the web page contains an object from a victim domain, which triggers a DNS request. 

(2) the recursive DNS resolver receives a DNS request and sends a DNS request to the target name server.

(3) the attacker then sends $2^{16}$ responses with spoofed source IP (of the name server); each response is a \texttt{referral}, mapping the name server {\sf ns.foo.com} to 6.6.6.6, an IP address controlled by the attacker.

(4) the response containing the correct TXID is accepted, cached and sent to the client.

(5) authentic DNS response is ignored, since there is no matching pending request.

Following to Kaminsky's attack, additional challenge-response mechanisms were proposed to increase the entropy in DNS requests (RFC 5452). 
The most popular mechanism, supported by most resolvers, is source port randomisation (SPR). SPR in tandem with TXID, produce a search space of $2^{32}$ and were believed to provide sufficient protection against poisoning by off-path adversaries, since this requires the attacker to not only guess the TXID but also the source port correctly. 
This reduced the motivation for deployment of the cryptographic defense against poisoning, DNSSEC (RFCs 4033-4035). 
\ignore{
Although DNSSEC was proposed and standardised more than 15 years ago, its deployment is very slow, and most domains and DNS resolvers do not support DNSSEC; (about 2\%) of the domains are signed
and about 1\% of the resolvers perform DNSSEC validation, see~\cite{olafur:nat:res} for more details.

However, we caution against this conclusion. The DNS protocol does not operate in an isolated (idealised) environment, but concurrently, and in parallel, to other systems and protocols, which can have an adverse impact on the DNS security, and the assumptions upon which the security is established can be invalidated.
}
We found different techniques, \cite{DBLP:journals/corr/abs-1205-4011,DBLP:conf/esorics/HerzbergS12}, to circumvent the popular defenses against poisoning by off-path attackers. 
In this work we show a simple technique from~\cite{DBLP:conf/esorics/HerzbergS12}, which applies to common network scenarios, where the DNS resolvers are located behind a NAT device (as in Figure~\ref{fig:ourmodel}), and uses side-channels for ports' prediction. 
\subsection{Vulnerability of Resolvers Behind NAT}
Network Address Translation (NAT) devices that assign predictable ports to outgoing packets can expose resolvers to cache poisoning attack. We show that even NAT devices that sufficiently randomise outgoing ports may expose resolvers to attacks.
\paragraph{Per-Destination NAT}
NAT devices modify ports in outgoing packets, and maintain mappings between internal and external ports.

We focus on the common {\em per-destination} NAT; see \cite{DBLP:conf/esorics/HerzbergS12} for other NAT devices. A per-destination NAT operates as follows: for a tuple defined by \url{<src-IP:src-port, dst-IP:dst-port,protocol>}, it selects the first port at random, and subsequent ports are increased sequentially (for that tuple). 

\paragraph{Predict-then-Poison Attack}
The off-path attacker, Oscar, controls a {\em zombie}, i.e., non-privileged malware, that runs on a client host in the LAN. 

\begin{figure}[!ht]
\centering
	\begin{center}
			    \epsfig{scale=0.42,file=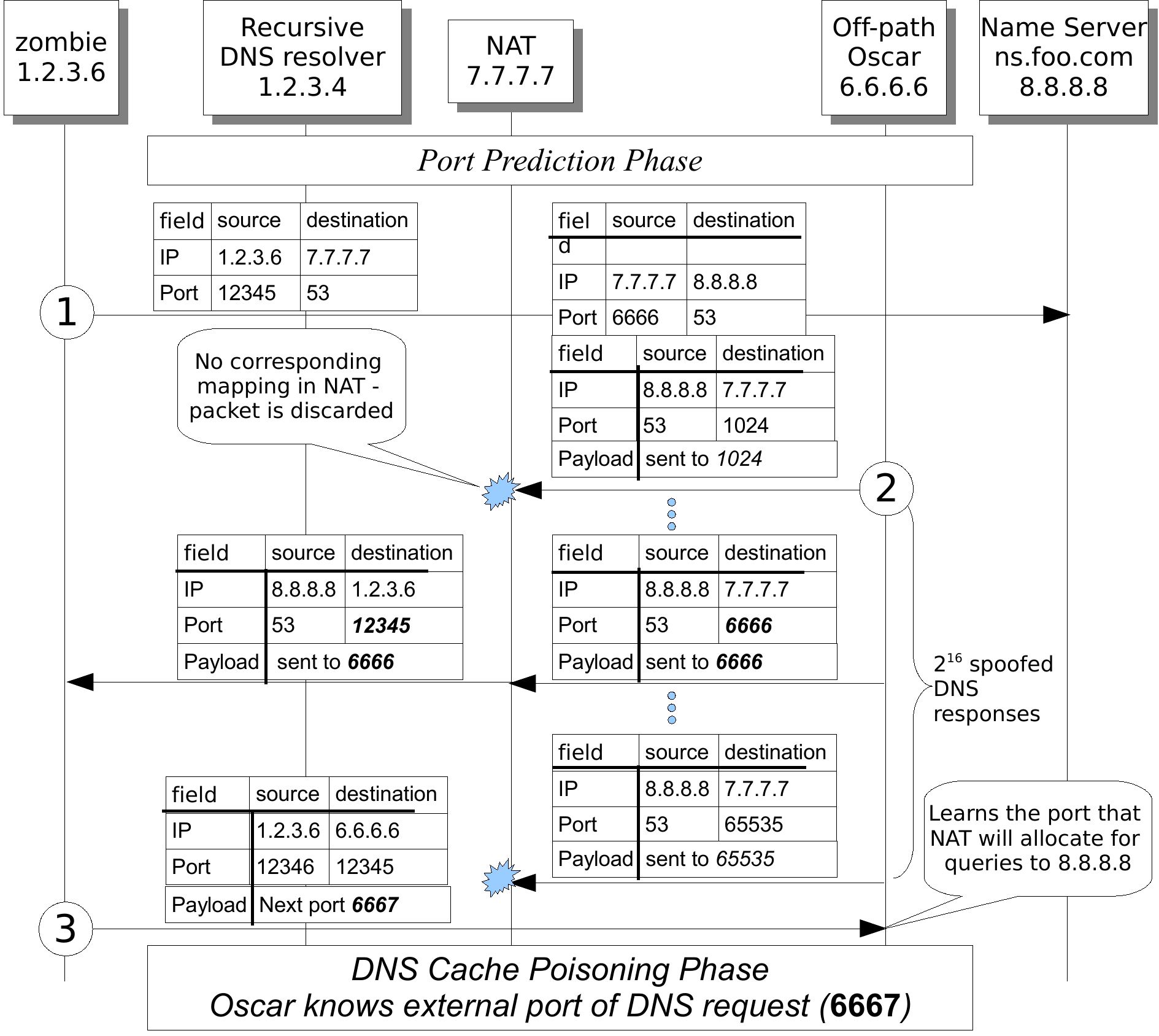}
	\end{center}	
\caption{Predict-then-Poison DNS Attack Assuming Per-Destination NAT.}
 \label{p:p}
\end{figure}

The first phase of the attack is {\em port prediction} during which the zombie and Oscar collaborate to expose the port that will be allocated to the DNS request of the DNS resolver. Then during the second phase, Kaminsky's attack is launched. We present only the {\em predict} phase of the attack, which steps are illustrated in Figure~\ref{p:p}; the steps of Kaminsky's cache poisoning attack are in Figure~\ref{fig:kaminsky}.

 (1) The zombie sends a packet to create a mapping in the NAT table;
  in the example in Figure~\ref{p:p} we assume arbitrarily that port 6666 was selected. 
  
 (2) Then, Oscar at address 6.6.6.6 sends $2^{16}$ packets with a spoofed source IP of 8.8.8.8, s.t., each is sent to a different port of the NAT and each contains the destination port in payload; only the packet with port 6666 arrives at the zombie. 
 
 (3) The zombie increments this port by 1, in our example the result is~6667, and sends it to Oscar; this is the external port that will be allocated by the NAT to the subsequent DNS request of the resolver to the victim name server. 
 
 This phase allows to bypass the SPR defense. 
 


%% file: TCP.tex
\section{TCP Injections}\label{sc:tcp}
The recent off-path TCP injection attacks operate in two phases:
\newline {\sf $\diamond$ Learn Connection 4-Tuple.} \mal, the off-path attacker, learns the four parameters of a TCP connection between a client and a server, that is, their respective IP addresses and ports. 
\newline {\sf $\diamond$ Learn Sequence Number(s).} \mal\ learns the current sequence number, for packets sent from the server to the client. In some attacks, \mal\ also learns the sequence number for packets from the client to the server. In this phase \mal\ bypasses the initial sequence number randomisation defense, which is implemented in all modern TCP/IP stacks.
 
Table \ref{Tbl:blocks} surveys the techniques used in recent injection attacks for the two phases and their requirements (in parentheses).
In the reminder of this section we present a simple implementation for each phase. 

\begin{table*}
\centering
{\scriptsize
\begin{tabular}{ c || c | c |}
                                 & Learn Connection 4-tuple         & Learn Sequence Numbers           \\ \hline \hline
                                     
        klm~\cite{klm:phrack:07} & $\begin{array}{c}\text{Active probing for connection} \\ \text{(Windows client, no firewall)}\end{array}$ & $\begin{array}{c}\text{Side channel} \\ \text{(Windows client)}\end{array}$\\ \hline

    $\begin{array}{c} \text{Qian} \\ \text{et al.~\cite{snptcp,CCS12:tcp}} \end{array}$  & $\begin{array}{c} \text{Monitor connections,}\\ \text{e.g., with {\sf netstat}} \\ \text{(Malware)}\end{array}$& $\begin{array}{c} \text{Read client system-counters,} \\ \text{(Malware; in~\cite{snptcp} also seq. \# checking firewall)}\end{array}$ \\ \hline
    	
  $\begin{array}{c}  \text{Gilad and}  \\ \text{Herzberg~\cite{woottcp}}  \end{array}$  
     & $\begin{array}{c} \text{Establish connection, exploit} \\ \text{sequential port allocation impl.} \\ \text{(Puppet, Windows client)}\end{array}$ & $\begin{array}{c}\text{Side channel} \\ \text{(Puppet, Windows client)}\end{array}$ \\ \hline
        
    $\begin{array}{c}  \text{Gilad and}  \\ \text{Herzberg~\cite{wwwtcp}}  \end{array}$                    & $\begin{array}{c}\text{Establish connection,}\\ \text{client port de-randomization} \\ \text{(Puppet, client behind firewall)}\end{array}$ & $\begin{array}{c}\text{Exploit browser behavior,} \\ \text{(Puppet, no TLS/SSL)}\end{array}$ \\ \hline
    
\end{tabular}
}
\caption{Off-Path TCP Injection Attacks: Building Blocks. In parentheses: requirements.}
\label{Tbl:blocks}
\end{table*}

\subsection{Learn Connection 4-Tuple} \label{establishconn}
In order to launch an injection attack, \mal\ must first identify a TCP connection between the victim client and~server.

We describe the method of~\cite{woottcp}, which uses a puppet (script restricted by browser sandbox) running on the client machine to {\em open} such a connection. Since \mal\ chooses the server, the server's IP address and port are known. To find the client's IP, the puppet sends a request to \mal's site; this request contains the client's IP address. 

The final challenge of this phase is to detect the client port. Many clients, in particular, those running Windows, assign ports to connections {\em sequentially}. The attack of~\cite{woottcp} uses the puppet to open a connection to \mal's remote site before and after opening the connection to the victim server; sequential port assignment allows \mal\ to learn the client's port: \mal\ observes $p_1$ and $p_2$, the client ports used in the connection to his sites. If $p_2 = p_1 + 2$, then he identifies that the connection to the server is via port $p_1 + 1$ (otherwise \mal\ restarts the attack). 

\subsection{Learn Sequence Numbers}
The next step after identifying the victim-connection, is learning one or both connection's sequence numbers. Observing the sequence numbers directly from traffic requires an on-path attacker (eavesdropping capability); off-path attacks use different methods to infer the sequence numbers. We focus on the technique of~\cite{wwwtcp} which exploits an {\em under-specification of HTTP} to learn the client's sequence number. 

{\sc Background.} As of HTTP~1.1, clients can send multiple requests to the same web-server in {\em pipeline} over a single (`persistent') HTTP connection. In order to allow browsers to match between each response and the corresponding request, the server sends the responses exactly in the order in which it received the requests. The browser (client) keeps a FIFO queue of pending HTTP requests for each connection, and handles them one by one, as follows. To handle a request, the browser reads the bytes in the TCP connection's receive-buffer (when they become available). The browser expects to find the matching response in the beginning of TCP's receive-buffer and parses the response.

The HTTP standard does not specify what the browser should do when the receive-buffer contains data which is {\em not} a valid (`parsable') HTTP response. Many modern browsers (including Chrome, Firefox and Internet Explorer) handle this situation as follows: the browsers treat {\em all available data} in the receive-buffer as payload of a response with the following `default'  HTTP header:

\begin{verbatim}
HTTP/1.1 200 OK
Content-Type: text/html; charset=us-ascii
Content-Length: available-data-size
\end{verbatim}

The browsers return this `response' to the requesting module, normally, the rendering engine or a script/applet.

\begin{figure}
  \begin{center}
    \includegraphics[width=0.52\textwidth]{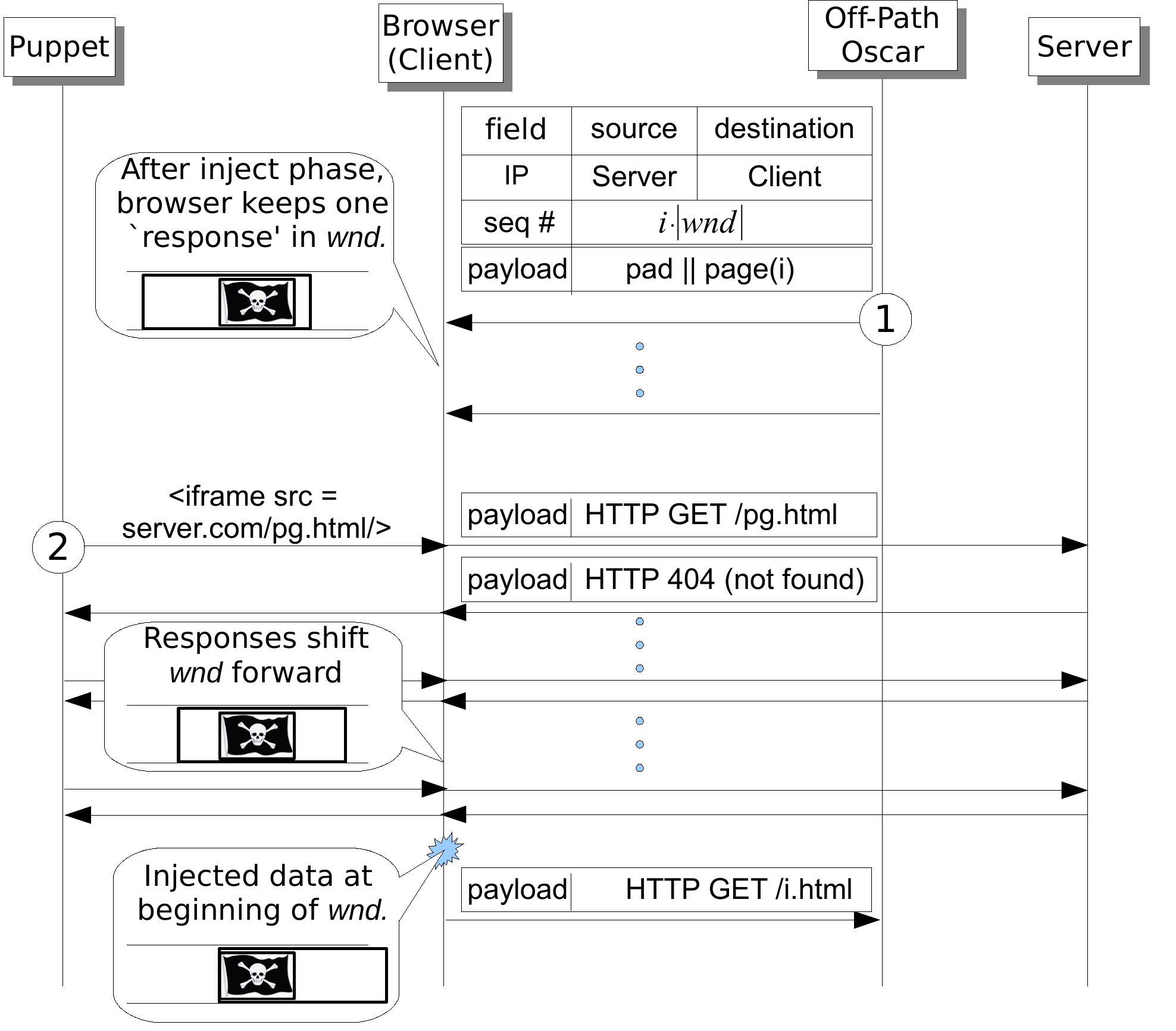}
  \end{center}
	\vspace{-7pt}
  \caption{Sequence Number Learning Technique.}
    \label{fig:injectandobserve}
	\vspace{-10pt}
\end{figure}

{\sc Sequence Number Learning.} The learning phase has two steps: {\em Inject} and {\em Observe}, see Figure~\ref{fig:injectandobserve}. In the inject step, \mal\ injects data into the stream of HTTP responses that the server sends to the client. This data is read in the observe step, which allows \mal\ to determine the server's sequence number.

{\sf (1) Inject step.} Let \wnd\ denote the browser's receive-buffer for the connection and $\wndsize$ denote its size.
In order to inject the data, \mal\ sends to the browser $\frac{2^{32}}{\wndsize}$ packets, spoofed to appear to be from the server (on its victim-connection with the client). The $i^{th}$ packet has server sequence number $i$~$\cdot$~$\wndsize$; since the sequence number field is 32-bits long, exactly {\em one} of these packets has a `valid' sequence number, which falls within \wnd; all the other packets are discarded by the client. Each of \mal's packets contains as payload $\textit{page}(i)$ which is a simple web-page defined as follows:
\begin{verbatim}
<HTML><BODY>
<iframe src = "oscar.com/i.html" />
</BODY></HTML>
\end{verbatim}

{\sf (2) Observe step.} In this step, the puppet makes prevalent requests to the server, until it reaches the data injected by \mal\ in the previous step. Each server-response that arrives at the client shifts \wnd\ forward; after several such responses arrive, there is no gap of unreceived bytes between the injected data and the beginning of \wnd. Then, the browser reads the injected-response, assuming that it corresponds to the request. 

The last server-response usually overwrites the beginning of the injected data, therefore, the injected `response' will usually be corrupt. However, as noted above, many browsers handle the injected data as payload `wrapped' with a default header. When the browser renders \textit{page(i)}, it tries to retrieve~\textit{i.html} from \mal's web-site (see Figure~\ref{fig:injectandobserve}); providing to \mal\ the value of $i$. In order to keep \textit{page(i)} intact despite the overwrite, when \mal\ sends \textit{page(i)} in the inject step, he prepends to it an easily removable pad. The value of $i$ allows \mal\ to compute the next server sequence number that the client expects. 

%% file: Exploits.tex
\section{Exploiting Injection/Poisoning}
To conclude our discussion of off-path TCP injection and DNS poisoning attacks, we briefly discuss some potential exploits. 

Exploiting {\em DNS poisoning} is straightforward, e.g., see \cite{kaminsky:dns}.  Both users and programs use DNS extensively to resolve domain names - to obtain the IP address and communicate with a server, or to obtain other DNS-indexed resource, often security related , e.g., SPF policies or blacklists. DNS poisoning allows circumvention of these mechanisms, e.g., `hijacking' of connection requests meant for legitimate servers.  In particular, `hijacking' can allow {\em phishing}, where a user thinks he interacts with a trusted site, while he actually deals with a fake site (exposing credentials, installing malware, etc.). Furthermore, the malicious mapping is kept for the time (TTL) specified in the record, and in the common case of a single cache used by many clients, the poisoning impacts all of them. 

Exploiting {\em TCP injections} is more challenging, since TCP is a transport protocol and does not involve caching. However, 
in common scenarios,  TCP injections can allow critical exploits. In particular, TCP injections suffice to {\em circumvent Same Origin Policy}, hijack `cookies' and cause execution of malicious scripts (XSS). Furthermore, to cause long-term impact similar to DNS poisoning, attackers can exploit caching of objects by web caches.  By crafting the HTTP headers of his response, \mal\ can cache the spoofed page and specify that it will be out-dated only after a long time; furthermore, when using a web-cache, the impact can be over many users. See details in  \cite{wwwtcp}.

%% file: conclusions.tex
\section{Conclusions}
The vulnerabilities presented in this work show that relying on challenge-response mechanisms against off-path attackers can often be circumvented. Specifically, the techniques discussed in this work allow off-path attackers to circumvent the main challenge-response defenses: {\em source port randomisation} and {\em initial sequence number randomisation}. 

Our message is that defenses should be designed and analysed against the strongest MitM attackers. In particular, to prevent these, and other unforeseen, vulnerabilities we recommend deployment of systematic, cryptographic defenses for TCP and DNS.